%% file: main.tex
\documentclass[sigconf,natbib=true,screen=true]{acmart}

\AtBeginDocument{%
  \providecommand\BibTeX{{%
    \normalfont B\kern-0.5em{\scshape i\kern-0.25em b}\kern-0.8em\TeX}}}

\setcopyright{acmlicensed}
\copyrightyear{2018}
\acmYear{2018}
\acmDOI{XXXXXXX.XXXXXXX}

\acmConference[Conference acronym 'XX]{Make sure to enter the correct
  conference title from your rights confirmation email}{June 03--05,
  2018}{Woodstock, NY}
\acmISBN{978-1-4503-XXXX-X/18/06}

\input{preamble/packages}

\input{preamble/definitions}

\begin{document}

\title[Can We Use Large Language Models to Fill Relevance Judgment Holes?]{Can We Use Large Language Models to \\ Fill Relevance Judgment Holes?}

\include{preamble/authors}

\begin{abstract}
Incomplete relevance judgments limit the re-usability of test collections. When new systems are compared against previous systems used to build the pool of judged documents, they often do so at a disadvantage due to the ``\textit{holes}'' in test collection (i.e., pockets of un-assessed documents returned by the new system). In this paper, we take initial steps towards extending existing test collections by employing Large Language Models (LLM) to fill the holes by leveraging and grounding the method using existing human judgments. We explore this problem in the context of \textit{Conversational Search} using \textit{TREC iKAT}, where information needs are highly dynamic and the responses (and, the results retrieved) are much more varied (leaving bigger holes). 
While previous work has shown that automatic judgments from LLMs result in highly correlated rankings, we find substantially lower correlates when human plus automatic judgments are used (regardless of LLM, one/two/few shot, or fine-tuned). We further find that, depending on the LLM employed, new runs will be highly favored (or penalized), and this effect is magnified proportionally to the size of the holes. Instead, one should generate the LLM annotations on the whole document pool to achieve more consistent rankings with human-generated labels.
Future work is required to prompt engineering and fine-tuning LLMs to reflect and represent the human annotations, in order to ground and align the models, such that they are more fit for purpose.

\end{abstract}


\maketitle

\input{sections/intro}

\input{sections/method}
\input{sections/results}

\input{sections/conclusion}

\bibliographystyle{ACM-Reference-Format}
\bibliography{sample-base}

\end{document}

%% file: preamble/packages.tex
\usepackage{color}
\usepackage{flushend}
\usepackage{graphicx}
\usepackage{booktabs}
\usepackage[htt]{hyphenat}
\usepackage[]{xcolor}
\usepackage{tabularx}
\usepackage{booktabs}
\usepackage{multirow}
\usepackage[inline]{enumitem}
\usepackage{xspace}
\usepackage{acronym}
\usepackage{arydshln}

%% file: preamble/definitions.tex
\definecolor{midnightgreen}{rgb}{0.0, 0.29, 0.33}

\acrodef{CSA}{Conversational Search Agent}
\acrodef{PTKB}{Personal Text Knowledge Base}
\acrodef{TREC}{TExt Retrieval Conference}
\acrodef{iKAT}{Interactive Knowledge Assistance Track}
\acrodef{CAsT}{Conversational Assistance Track}
\acrodef{NIST}{National Institute of Standards and Technology}
\acrodef{LLM}{Large Language Model}

\acrodef{QPP}{query performance prediction}
\acrodef{IR}{information retrieval}
\acrodef{NLP}{natural language processing}
\acrodef{PEFT}{parameter-efficient fine-tuning}
\acrodef{ICL}{in-context learning}
\acrodef{LoRA}{low-rank adaptation}

%% file: preamble/authors.tex
\author{Zahra Abbasiantaeb}
\orcid{}
\affiliation{%
  \institution{University of Amsterdam}
  \city{Amsterdam}
  \country{The Netherlands}
}
\email{z.abbasiantaeb@uva.nl}

\author{Chuan Meng}
\orcid{0000-0002-1434-7596}
\affiliation{%
  \institution{University of Amsterdam}
  \city{Amsterdam}
  \country{The Netherlands}
}
\email{c.meng@uva.nl}

\author{Leif Azzopardi}
\orcid{0000-0002-6900-0557}
\affiliation{%
  \institution{University of Strathclyde}
  \city{Glasgow}
  \country{Scotland, UK}
}
\email{leif.azzopardi@strath.ac.uk}

\author{Mohammad Aliannejadi}
\orcid{0000-0002-9447-4172}
\affiliation{
  \institution{University of Amsterdam}
  \city{Amsterdam}
  \country{The Netherlands}
}
\email{m.aliannejadi@uva.nl}

\renewcommand{\shortauthors}{Zahra Abbasiantaeb, Chuan Meng, Leif Azzopardi, and Mohammad Aliannejadi}

%% file: sections/intro.tex
\section{Introduction}

Building reusable test collections in a cost-efficient manner to evaluate current and future systems has been a long-standing challenge in the field of Information Retrieval (IR)~\cite{cleverdon1967}. 
The predominant strategy for creating such collections has been through the use of pooling~\cite{sparck-jones1975,voorhees2005trec} -- where a subset of documents, taken from various systems, is assessed for relevance. 
This is a compromise away from the ``ideal test collection'' with complete relevance assessments which is infeasible and impractical. 
And while the pooling strategy is fairly robust~\cite{cormack1998efficient_test_collections, buckley2004incomplete_evaluation, sanderson2005ir_evaluation,lu2016effect}, it leads to various evaluation biases (e.g.,~\cite{baillie2007assessed_metric, buckley2004incomplete_evaluation}) where systems that did not contribute to the pool, can be significantly disadvantaged\footnote{
This is because documents that haven't been judged are considered irrelevant. 
Therefore, the fewer judged/assessed documents returned in a ranking, the lower the retrieval performance ceiling~\cite{baillie2008assessed_metric}.}. 
However, the fewer the judgments required to compare systems the cheaper the test collection.

While researchers have tried to address these trade-offs in various ways, either by proposing new metrics and methodologies for compensating for the un-assessed documents (e.g., ~\cite{buckley2004incomplete_evaluation,aslam2007inferred_ap,baillie2008assessed_metric}), and/or developing new pooling and judgment strategies (e.g., ~\cite{carterette2006minimal_test_collections,lipani2017pooling_strategies,moffat2007strategic_judgments}), ``holes'' in the pools still remain~\cite{vorhees2022holes,macavaney2023holes}.

However, with the advances in the development of powerful Large Language Models (LLMs) and other neural-based models, new opportunities arise for building scalable, robust, and reusable test collections at a lower cost. \acp{LLM} offers the possibility to:
(1) assess large volumes of documents reasonably cheaply, especially compared to human judgments, 
(2) do so in a consistent and independent but potentially biased manner, 
(3) if the LLM and prompt are fixed and shared, then judgments can be collected at different times under the same conditions, and,
(4) typically at a higher quality than ``typical'' crowd workers.

\input{prompts/prompt2}

Indeed, recent studies~\citep{meng2024query,khramtsova2024leveraging,thomas2023large,faggioli2023perspectives} have shown the effectiveness of using \acp{LLM} to automatically generate relevance judgments in the scenario of ad-hoc search. 
These works demonstrate that LLM-based judgments exhibit a high correlation with human judgments.
\citet{thomas2023large,faggioli2023perspectives} prompted commercial \acp{LLM} (e.g., ChatGPT, GPT-3.5/4) to generate relevance judgments; however, commercial \acp{LLM} come with limitations like non-reproducibility, non-deterministic outputs and potential data leakage between pre-training and evaluation data, impeding their utility in scientific research~\citep{pradeep2023rankzephyr}.
\citet{macavaney2023holes} and \citet{khramtsova2024leveraging} prompted
an open-source (OS) \acp{LLM}, Flan-T5~\citep{chung2022scaling}, for generating relevance judgments. While OS \acp{LLM} are less effective, they do offer the potential for the development of reproducible and reusable test collections at scale. 
This led to efforts by
\citet{meng2024query}, who fine-tuned a leading open-source \ac{LLM}, LLaMA~\citep{touvron2023llama} using \ac{PEFT}~\citep{dettmers2023qlora} to better condition the \ac{LLM} for performing the task of assigning relevance judgments.
They found that fine-tuned LLaMA leads to better agreement with human annotators than ChatGPT~\citep{faggioli2023perspectives}. 
Further suggesting that more complete test collections could be produced, with high quality, for lower cost. 

While this prospect is very appealing, it is fraught with new, unexplored challenges. 
Of interest, in this work is the notion of grounding. 
Training systems on the judgments of \acp{LLM}, and then evaluating those systems on subsequent test collections, based on judgments from \acp{LLM}, creates a potentially dangerous cycle that may amplify and re-enforce existing biases inherent in \acp{LLM}. 
Grounding the \acp{LLM} based judgment given human judgments provides a mechanism to condition the \acp{LLM} to be more aligned with human annotators -- as a means to avoiding the hypothesized AI doom loop~\cite{peterson2024ai}. 
To this end,  \citet{macavaney2023holes} focus on a setting wherein the \ac{LLM} is given one relevant example to help ground the subsequent judgments. 
In this paper, we draw up this direction in the context of Conversational Search (CS) and aim to build/augment test collections with grounded \ac{LLM} based judgments.

Conversational search (CS) is defined as responding to the user's information needs in the context of the conversation~\citep{radlinski2017cs,azzopardi2018cs,meng2023query,meng2023system}. 
In CS the user's information need depends on the query, the context of the conversation, and the user's personal preferences~\cite{aliannejadi2024trec}. 
This results in highly dynamic, non-linear conversational trajectories -- where a user's information need could be answered quite differently depending on the system's interpretation because the needs are evolving and change in response to the information presented.
Therefore, to service such dynamic information needs, systems are likely to retrieve a much greater array of documents -- and this could lead to many more un-judged documents (bigger holes) when evaluating new/future systems -- severely limiting a test collections re-usability~\cite{abbasian2024generate}. So, in this work, we explore whether LLMs can be used to augment and extend CS test collections which are grounded by human annotations, in order to evaluate new, future systems.

\input{Tables/dist-ft-dataset}

In this paper, we leverage both commercial and open-source \acp{LLM} in a zero- and few-shot, as well as fine-tuning manners, to automatically generate relevance judgments in the CS scenario. 
Specifically, for commercial \acp{LLM}, we use the ChatGPT model with different prompts. We try one-shot, two-shot, and zero-shot prompts. 
For open-source \acp{LLM}, we consider three setups: 
(1) we directly use the LLaMA checkpoint released by \citet{meng2024query}, which has undergone fine-tuning based on human-labeled relevance judgments on MS-MARCO, 
(2) we directly prompt LLaMA-3~\citep{llama3modelcard} in a zero-shot way, and
(3) we first use partial human-labeled relevance judgments in a CS dataset to fine-tune LLaMA-3 and then test it using the rest of the relevance judgments in the dataset.

We use the TREC iKAT~\cite{aliannejadi2024trec} dataset which is a personalized CS benchmark. 
In our experiments, we re-create the relevance judgments of the TREC iKAT benchmark using various prompts and techniques.
We compare the generated judgments with the official TREC iKAT relevance labels in terms of various metrics. In particular, we are interested in answering the following research questions:
\begin{enumerate}[label=\textbf{RQ\arabic*}]
    \item How do different \acp{LLM} compare in predicting relevance judgments in conversational search? \label{RQ1}    
    \item How do LLM-generated assessments compare to human-generated assessments in both absolute label prediction and relative ranking of retrieval models? \label{RQ2}
    \item How are new models with different levels of holes ranked using LLM-generated assessments? Can we rely on LLM-generated labels to compare a new model with existing models? \label{RQ3}
\end{enumerate}

To answer the RQs, we conduct a set of experiments where for \ref{RQ1}, we create a training and test set of relevance labels and compare LLaMA-1, LLaMA-3, and ChatGPT in zero-shot, few-shot, and fine-tuning settings. 
For \ref{RQ2}, we use ChatGPT to generate relevance labels on the official TREC iKAT 2023 pool and use it to rank the official TREC runs. To answer \ref{RQ3}, we conduct multiple experiments where at each experiment, we remove the relevance labels of each run from the pool\footnote{We do not remove the documents that are in common with other runs.}, simulating the case where a new model is being assessed using the original pool --- every time one of the models is removed, mimicking the case where that model is not included in the original pool. We then generate the relevance labels using ChatGPT and use those labels to assess the new model.

{Our results show that ranking of the retrieval models using human- and LLM-generated are highly correlated, although the LLM-generated judgments and human judgments have a low agreement in binary- and graded-level, in line with the findings of \citet{faggioli2023perspectives}. 
In addition, the correlation of different IR metrics converges as we add more retrieval systems to the comparison pool.
This means that only LLM-generated pools are reliable for comparing different retrieval models in CS.
We show that by fine-tuning the open-source LLMs we can achieve a higher agreement between LLM-generated judgments and human judgments. 
However, higher agreement in terms of binary and graded judgments does not necessarily result in a higher ranking correlation.  
Moreover, we show that in case of adding a new retrieval model, filling the hole with a zero-shot LLaMA model results in less significant shifts
in the ranking of the corresponding retrieval model compared to using one-shot ChatGPT, perhaps due to higher agreement of the ratings, and because the one-shot ChatGPT model is biased to predict higher relevance scores. }

%% file: prompts/prompt2.tex
\begin{table}[]
    \centering
    \caption{The template of the prompts designed for relevance judgment. The orange rows belong to the one-shot prompt, the blue rows belong to two-shot prompt. The black lines are in all prompts including zero-shot, one-shot and few-shot.}
    \begin{tabular}{|p{8cm}|}
    \hline
    \textbf{The prompts used for relevance judgment} \\ \hline
    \\
    \textbf{Instruction:} You are a search quality rater evaluating the relevance of web pages. \\
    Given the persona of the user, user query, and a web page, you must provide a score on an integer scale of 0 to 4 to indicate to what extent the given document meets the information needs of the user. The scores have the following meanings: \\ \\
     \hdashline \\
    0: fails to meet  \\
    1: slightly meets  \\
    2: moderately meets  \\
    3: highly meets  \\
    4: fully meets  \\
    \\
    User persona: \{ ptkb \} \\
    Query: \{ utterance \} \\
    \\

    \textcolor{orange}{document 1: \{ canonical response \} }\\
    \textcolor{orange}{Score: \{ 4 \} }\\
    \\
    
    \textcolor{blue}{document 1: \{ passage 1 \} }\\
    \textcolor{blue}{Score: \{ score 1 \} }\\
    \\
    \textcolor{blue}{document 2: \{ passage 2 \} }\\
    \textcolor{blue}{Score: \{ score 2 \}} \\
    \\
    
    document : \{ document \} \\
    Score: \\
        \\
    Please only generate an int score between 0 to 4 to say to what extent the document is relevant to the user question. Score lower than 2 means the document is not relevant. \\
   
\hline
    \end{tabular}
    \label{tab:two-shot-prompt}
\end{table}

%% file: Tables/dist-ft-dataset.tex
\begin{table}[t]
    \caption{Distribution of the relevance scores in train, test, and validation sets.}
    \centering
    \begin{tabular}{lcccccc}
\toprule
 & 0 & 1 & 2 & 3 & 4 & Total\\ 
\midrule
Train & 1551 & 217 & 1289 & 661 & 134 & 3852\\ 
Test & 389 & 60 & 295 & 150 & 23 & 917 \\  
Validation & 299 & 48 & 219 & 121 & 22 & 709 \\  
\bottomrule
    \end{tabular}
    \label{dist-ft-dataset}
\end{table}

%% file: sections/method.tex
\section{Methodology}
\label{sec:methodology}

\subsubsection*{\bf The choice of LLMs} 
As a commercial closed-source LLM, we consider the ChatGPT (gpt-3.5-turbo-0125) model with values of 0 and 1 for the \textit{temperature} and \textit{top\_p=1}. The value of 0 for temperature means that the model outputs the tokens with the highest probability and has no randomness.
For open-source LLMs, we consider the following three setups:
(1) we directly use the LLaMA (7B) checkpoint released by \citet{meng2024query},\footnote{\url{https://github.com/ChuanMeng/QPP-GenRE}} which has undergone fine-tuning using the human-labeled relevance judgments from the development set of MS~MARCO~V1~\citep{bajaj2016ms};
(2) we directly prompt LLaMA-3 (8B)~\citep{llama3modelcard} in a zero-shot way; specifically, besides the original version of LLaMA-3 (8B), we also consider its instruction-tuned version (Llama-3-inst);
(3) we first use partial human-labeled relevance judgments in a CS dataset to fine-tune the two variants LLaMA-3, and then test them using the rest of the relevance judgments in the dataset.

\input{Tables/kappa-test}

\begin{figure}[!b]
    \centering
    \includegraphics[width=0.5\textwidth]{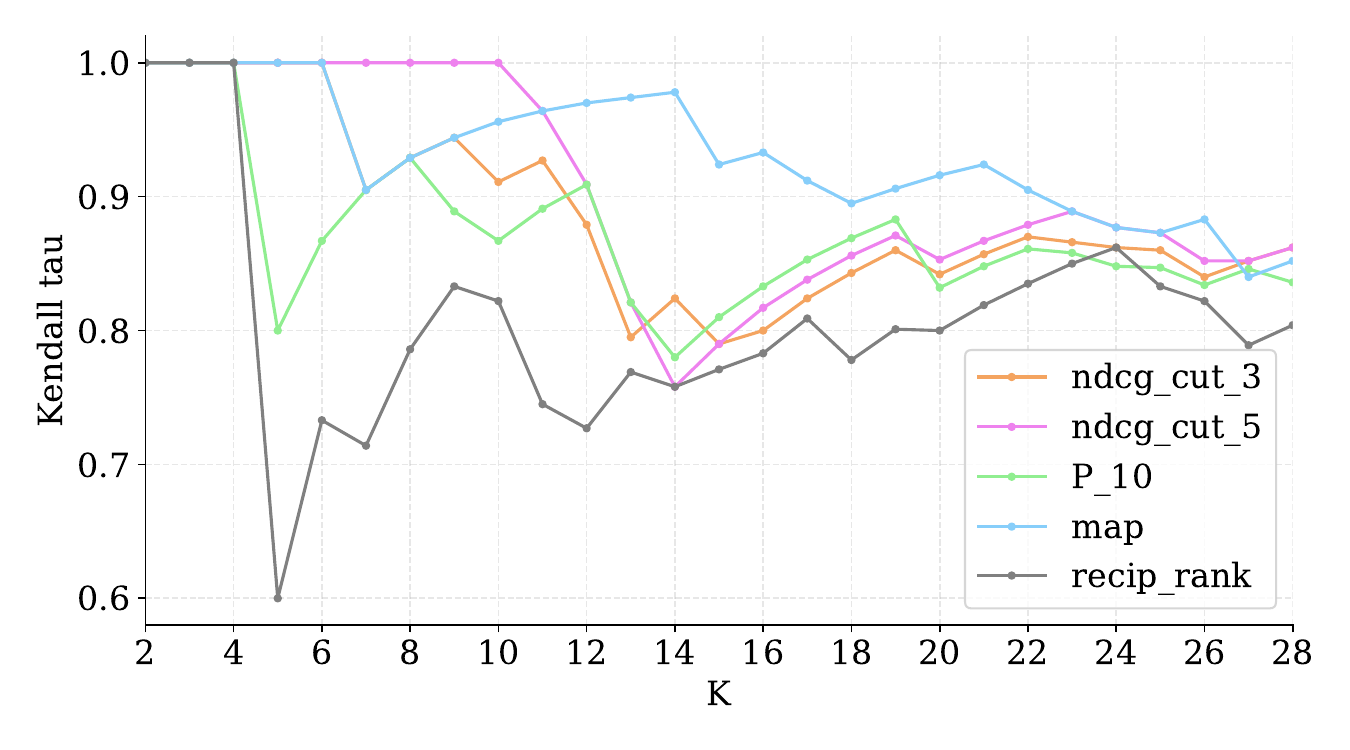}
    \caption{Ranking correlation between human- and LLM-generated pools using the $K$ best-performing models from TREC iKAT 2023 runs according to the human-generated pool. The ChatGPT one-shot ($tmp=0$) model is used for generating the LLM-based pool. }
    \label{fig:kendall@k}
\end{figure}

\input{Tables/ranking-replacing-test-data}

\subsubsection*{\bf Dataset} We use the TREC iKAT 2023~\cite{aliannejadi2024trec} benchmark in our experiments. We use the same grading scale used for benchmarking the TREC iKAT 2023 in our experiments. The relevance of each query--document pair is assessed and represented with a score in the range of 0--4. Using the ChatGPT model, we judge the relevance of all query--document pairs from the TREC iKAT 2023 collection. 
For fine-tuning the Llama-3 model~\citep{llama3modelcard}, we divide the TREC iKAT 2023 benchmark into train, test, and validation sets. First, we randomly remove 20680 non-relevant documents ($score<2$) from the existing pool to ensure that the number of relevant ($score>=2$) and non-relevant documents for each query are the same. 
Second, we randomly split the documents for each query between train, test, and validation sets. We keep the portion of train, test, and validation sets as 70\%, 15\%, and 15\%. The train, test, and validation sets include 3852, 917, and 709 query-passage pairs, respectively. 
We ensure that all user queries appear in the training set and have at least one positive and one negative document. Statistics of the train, test, and validation sets are shown in Table~\ref{tab:dist-ft-dataset}.

\subsubsection*{\bf Retrieval models} To assess the correlation of the generated pools, we use the baselines and runs submitted to TREC iKAT 2023. There are in total 28 baselines and runs submitted to the TREC iKAT 2023. We use the output of retrieval for these runs and baselines which are released by the organizers. 

\subsubsection*{\bf Metrics}  To assess the ranking-level performance of our proposed models for relevance judgment, we rank the retrieval models two times, once based on their performance using the main pool and second based on the generated pool. We compute and report Kendall's Tau, Spearman's Rho, and Rank-Boased Overlap (RBO) metrics to assess the correlation between the two rankings. To assess the agreement between proposed models for relevance judgment and humans, we report Cohen's Kappa at binary and graded levels. 
We convert the predicted graded scores (0-4) to a binary label by considering scores 2-4 as relevant and scores 0-1 as irrelevant. 

We compute the correlation with expert annotators on test set.

\subsubsection*{\bf ChatGPT prompt design}
For different experimental setups, we design a total of three different prompts, inspired by relevant work.
\begin{itemize}[nosep, leftmargin=*]
    \item We design a zero-shot prompt inspired by the prompt used in \citet{thomas2023large} which is shown in Table~\ref{tab:two-shot-prompt}.
    
    \item Our next prompt is a one-shot prompt which includes a relevant document with a relevance score of 4. This prompt is inspired by the prompt used by \citet{macavaney2023holes}. This prompt is shown in Table~\ref{tab:two-shot-prompt}. In this prompt, we use the canonical response for the corresponding user utterance from the TREC iKAT 2023 collection as the relevant (perfect) example with a relevance score of 4. The canonical responses are provided by the organizers and are supposed to be the best possible answer that can be given to the utterance at every point in the conversation~\cite{aliannejadi2024trec}. 

    \item The third prompt includes a relevant document and a non-relevant document. We randomly sample these documents from the TREC iKAT 2023 official relevance judgments. A document with a relevance score higher than or equal to 2 is selected as relevant and a document with a relevance score lower than 2 is randomly selected as non-relevant. This prompt is shown in Table~\ref{tab:two-shot-prompt}.

\end{itemize}

\subsubsection*{\bf LLaMA fine-tuning and prompt design}
For open-source LLMs, we consider the following setups:
\begin{itemize}[nosep,leftmargin=*]
    \item We use the prompt shown in Table~\ref{tab:two-shot-prompt} to prompt LLaMA-3.
    
    \item For fine-tuning LLaMA-3, we follow \citet{meng2024query} to fine-tune LLaMA-3 using a novel \ac{PEFT} method, 4-bit QLoRA~\citep{dettmers2023qlora}; the train, test, and validation data used for fine-tuning and inference over the model is explained above.
    
    \item Because the LLaMA model released by \citet{meng2024query} is only trained to generate binary relevance judgments given a query and a passage on MS MARCO, we leave out the user personal knowledge when we use the LLaMA model released by \citet{meng2024query}.
\end{itemize}

%% file: Tables/kappa-test.tex
\begin{table}[t]
    \centering
    \caption{Kappa Cohen's agreement between human and LLM labels on the test set at binary and graded levels.}
    \begin{tabular}{lcc}
\toprule
Hole Filling Model & Binary & Graded \\ 
\midrule
zero-shot ($tmp=1$)& 0.410 & 0.151 \\ 
zero-shot ($tmp=0$) & 0.489 & 0.117 \\ \midrule
one-shot ($tmp=1$)& 0.499 & 0.212 \\ 
one-shot ($tmp=0$) & 0.543 & 0.212 \\ \midrule
two-shot ($tmp=1$)& 0.329 & 0.134 \\ 
two-shot ($tmp=0$) & 0.454 & 0.184 \\ \midrule
Llama-3 zero-shot & 0.015 & -0.005 \\
Llama-3-inst zero-shot & 0.127 & 0.092 \\ \midrule
Llama-3-inst FT & \textbf{0.729} & \textbf{0.553} \\ 
Llama-3 FT & 0.687 & 0.527 \\  \midrule \midrule

Llama-1 (pre-trained)~\citep{meng2024query} & 0.386 & - \\
\bottomrule

    \end{tabular}
    \label{tab:kappa-test}
\end{table}

%% file: Tables/ranking-replacing-test-data.tex
\begin{table*}[t]
    \centering
    \caption{Comparison between the relative ranking of TREC iKAT 2023 runs using (1) subset of human-generated pools on the test set and (2) LLM-generated labels on the same test subset. The relative ranking is compared using Kendall's Tau ($\tau$), Spearman's Rho ($\rho$), and Rank-Biased Overlap (RBO) metrics. }
    \begin{tabular}{llcccccccc}
    \toprule
Metric & Hole Filling Model  & ndcg\_cut\_3 & ndcg\_cut\_5 & ndcg & P\_10 & recall\_10 & recall\_1000 & map & recip\_rank \\ \midrule
$\tau$ & zero-shot ($tmp=1$) & 0.788 & 0.772 & 0.820 & 0.857 & 0.709 & 0.884 & 0.746 & 0.815 \\ 
 & zero-shot ($tmp=0$) & \textbf{0.852} & \textbf{0.799} & \textbf{0.862} & 0.820 & 0.767 & 0.868 & 0.741 & 0.810 \\
& one-shot ($tmp=1$) & 0.836 & 0.794 & \textbf{0.862} & 0.820 & 0.804 & \textbf{0.931} & 0.783 & 0.794 \\ 
 & one-shot ($tmp=0$) & 0.783 & 0.778 & 0.831 & \textbf{0.915} &\textbf{ 0.841} & 0.921 & 0.772 & 0.847 \\
 & two-shot ($tmp=1$) & 0.810 & 0.762 & 0.810 & 0.841 & 0.794 & 0.894 & \textbf{0.794} & 0.873 \\ 
 & two-shot ($tmp=0$) & 0.767 & 0.746 & 0.772 & 0.884 & 0.810 & 0.884 & 0.735 & 0.804 \\ \cmidrule{2-10}
 & Llama-3 FT & 0.788 & 0.772 & 0.868 & 0.905 & 0.810 & 0.915 & 0.772 & \textbf{0.921} \\ 
 & Llama-3-inst FT & 0.762 & 0.751 & 0.820 & 0.894 & 0.825 & 0.894 & 0.741 & 0.847 \\ 
 & Llama-3 zero-shot & 0.614 & 0.571 & 0.778 & 0.630 & 0.614 & 0.820 & 0.640 & 0.735 \\
 & Llama-3-inst zero-shot & 0.550 & 0.534 & 0.788 & 0.624 & 0.550 & 0.868 & 0.635 & 0.614 \\ 
\bottomrule
$\rho$ & zero-shot($tmp=1$)  & 0.928 & 0.890 & 0.941 & 0.957 & 0.862 & 0.958 & 0.890 & 0.949 \\
 & zero-shot ($tmp=0$) & \textbf{0.957} & 0.912 & 0.962 & 0.943 & 0.907 & 0.971 & 0.882 & 0.946 \\
 & one-shot ($tmp=1$) & 0.954 & \textbf{0.927} &\textbf{ 0.963} & 0.946 & 0.926 & \textbf{0.986} & 0.910 & 0.933 \\ 
 & one-shot ($tmp=0$) & 0.919 & 0.915 & 0.952 & \textbf{0.986} & \textbf{0.944} & 0.984 & 0.900 & 0.955 \\
 & two-shot ($tmp=1$) & 0.932 & 0.892 & 0.934 & 0.952 & 0.930 & 0.976 & \textbf{0.916} & 0.967 \\
 & two-shot ($tmp=0$) & 0.910 & 0.870 & 0.914 & 0.973 & 0.929 & 0.966 & 0.869 & 0.938 \\ \cmidrule{2-10}
 & Llama-3 FT & 0.923 & 0.895 & 0.955 & 0.966 & 0.926 & 0.970 & 0.899 & \textbf{0.980} \\
 & Llama-3-inst FT & 0.909 & 0.886 & 0.921 & 0.963 & 0.928 & 0.968 & 0.864 & 0.946 \\
 & Llama-3 zero-shot & 0.783 & 0.751 & 0.924 & 0.821 & 0.775 & 0.944 & 0.801 & 0.888 \\
 & Llama-3-inst zero-shot & 0.706 & 0.708 & 0.924 & 0.773 & 0.706 & 0.963 & 0.806 & 0.772 \\ 
\bottomrule
RBO  & zero-shot ($tmp=1$) & 0.915 & 0.915 & 0.897 & 0.941 & 0.891 & 0.955 & 0.930 & 0.925 \\
 & zero-shot ($tmp=0$) & \textbf{0.945} & 0.934 & 0.911 & 0.906 & 0.912 & 0.945 & 0.920 & 0.917 \\ 
    & one-shot ($tmp=1$) & 0.933 & 0.924 & \textbf{0.956} & 0.943 & 0.924 & \textbf{0.979} & 0.930 & 0.933 \\
 & one-shot ($tmp=0$) & 0.917 & 0.924 & 0.899 & 0.948 & \textbf{0.945} & 0.973 & \textbf{0.932} & 0.929 \\
 & two-shot ($tmp=1$) & 0.932 & 0.930 & 0.900 & 0.941 & 0.925 & 0.955 & 0.931 & \textbf{0.948} \\
 & two-shot ($tmp=0$) & 0.910 & 0.904 & 0.887 & 0.942 & 0.933 & 0.954 & 0.918 & 0.911 \\ \cmidrule{2-10}
 & Llama-3 FT & 0.916 & 0.918 & 0.922 & \textbf{0.955} & 0.927 & 0.965 & 0.927 & 0.970 \\ 
 & Llama-3-inst FT & 0.913 & \textbf{0.938} & 0.913 & 0.933 & 0.944 & 0.951 & 0.914 & 0.941 \\
 & Llama-3 zero-shot & 0.838 & 0.835 & 0.865 & 0.807 & 0.864 & 0.899 & 0.859 & 0.852 \\
 & Llama-3-inst zero-shot & 0.762 & 0.761 & 0.913 & 0.834 & 0.775 & 0.924 & 0.801 & 0.775 \\ 
\bottomrule

    \end{tabular}    
    \label{tab:1-ranking-replacing-test-set}
\end{table*}

%% file: sections/results.tex
\section{Experiments \& Results}
In this section, we describe in detail the experiment we design to answer each research question, followed by the results we obtain by doing the experiments.

\input{Tables/confusion}

\input{Tables/kappa}

\input{Tables/one-shot-ranking}

\

\subsection{LLM relevance label comparison (RQ1)}
\subsubsection{\bf Experimental design}
In this section, we aim to answer our first research question, \textit{\ref{RQ1}: How do different LLMs compare in predicting relevance judgments in conversational search?}
To do so, as described in Section~\ref{sec:methodology}, we randomly sample the human-generated labels into the train, validation, and test sets and use the training data to fine-tune LLaMA-based models. We then compare the performance of the LLaMA-based models with the different ChatGPT-based models.

\subsubsection{\bf Results}
In Table~\ref{tab:kappa-test}, we report the agreement of our proposed models on the test set. The experiments reveal that by fine-tuning the Llama-3-inst model we can highly improve the agreement. As can be seen, the fine-tuned Llama-3-inst achieves the agreement of 0.729 on the binary level.

We use fine-tuned and zero-shot LLaMA to predict the test set. We create a small pool based on the predictions of the model on the test set. 
We sorted the the TREC iKAT 2023 runs based on their retrieval performance two times (1) using the LLM-generated assessments and (2) using the human-generated pool.
We compare these generated sorted lists of runs by computing the correlation between them. Table~\ref{tab:1-ranking-replacing-test-set} reports the result of the relative ranking performance of different LLMs, compared to human ranking.
Surprisingly, the LLaMA-3 model is not performing better than the ChatGPT model in this scenario while it has a higher agreement with human labels on the same test set. This could be due to the different labeling biases that the models have where ChatGPT labels could be more different from human labels in terms of absolute numbers, but when we compare different documents they are more similar relatively.

We report the binary- and graded-level confusion matrices for prediction of best LLaMA- and ChatGPT-based models over the test set in Tables~\ref{tab:confusion-binary} and \ref{tab:confusion-graded}, respectively. Additionally, we report the binary confusion matrix of the LLaMA-1 model which is fine-tuned on the MS MARCO dataset.
According to Table~\ref{tab:confusion-graded}, the fine-tuned Llama-3-inst model has a very lower tendency to assign scores 1 and 4 compared to the scores 0 and 2. The behavior of the LLaMA model is natural as the train data has less number of 1 and 4 labels compared to other labels according to Table~\ref{dist-ft-dataset}. We do not observe such bias in the one-shot ChatGPT model as this model is not fine-tuned on the train data. However, the one-shot ChatGPT model has predicted a large number of 4 labels compared to the LLaMA model. This bias could be the result of putting the canonical answer in the prompt as a positive example with a score of 4. 

Looking at the confusion matrices at the binary level, we observe that the one-shot ChatGPT model has more false positives compared to the fine-tuned LLaMA model. Giving one positive example in the prompt might cause this bias. 

The distribution of the false positive and false negative is approximately equal for the LLaMA model. This might be because of the fact that the number of the relevant and non-relevant passages in the training set of LLaMA are equal. 
Note that the LLaMA-1 model which is fine-tuned on the MS MARCO dataset has the lowest accuracy in terms of binary labels.

\subsection{LLM vs.\ human labels (RQ2)}
\subsubsection{\bf Experimental design}
Here, we aim to answer our second research question, \textit{\ref{RQ2}: How do LLM-generated assessments compare to human-generated assessments in both absolute label prediction and relative ranking of retrieval models?}
To do so, we regenerate all the relevance labels of the official TREC iKAT 2023 pool using the three prompts described in Section~\ref{sec:methodology}.
Inspired by \citet{faggioli2023perspectives} and \citet{macavaney2023holes} we aim to test the hypothetical case of having zero or one assessed passage for each query and rely on LLMs to generate the assessments on the whole pool.
In this experiment, we evaluate the models based on both the quality of individual predicted labels and the relative ranking of the models assessed with each of the generated labels, compared to human labels.

\subsubsection{\bf Results}
We report the agreement of proposed models with human labels over the complete pool of the TREC iKAT 2023 dataset in Table~\ref{tab:kappa-all}. 
According to the table, one-shot prompting the ChatGPT model has the highest agreement with human labels among zero-shot and two-shot prompting. 
Additionally, setting the temperature to 0 increases the agreement over all settings. 
A lower value for the temperature parameter means that the model has less randomness in generating the output while higher values mean that the model is more creative and has randomness in generation. 
Consequently, in relevance judgment, the temperature value of 0 increases both the binary and graded level agreement between the model and human assessors as the model selects the output with the highest probability. 
Our one-shot prompt has the highest agreement in terms of both binary and graded labels. The better performance of one-shot prompting compared to two-shot prompts indicates that (1) using the canonical response as a positive example is more useful and (2) using two positive and negative examples confuses the ChatGPT model. 
Additionally, we can conclude that NIST assessors\footnote{The TREC iKAT pool is judged by NIST assessors who are professional in relevance judgment.} consider the canonical response in their judgments.

We use the ChatGPT model in different settings and do the relevance judgment over all query--passage pairs of the TREC iKAT 2023 pool. 
As a result, we create several LLM-generated pools using each setting.
This way we can compare the quality of the LLM-generated vs.\ human-generated assessments.
More correlation between these rankings indicates that using LLM-generated assessments is as efficient as using human-generated assessments. 
Table~\ref{tab:1-shot-rank} shows the correlation between the relative ranking of TREC iKAT 2023 runs using different LLM-generated pools with the human-generated pool. Here we use 28 runs and baselines from TREC iKAT 2023. 
As can be seen, one-shot prompting the ChatGPT model significantly outperforms the other settings over all correlation metrics namely, Kendall's Tau, Spearman's Rho, and RBO.
Interestingly, we observe that using a temperature of 0 cannot outperform the case of using temperature of 1 in terms of all retrieval metrics. 
For example, one-shot prompting the ChatGPT model with the temperature of 1 achieves a higher correlation compared to using the temperature of 0 in terms of ndcg\_cut\_3 and ndcg\_cut\_5.

In Figure~\ref{fig:kendall@k}, we show the correlation between the relative ranking of LLM-generated and human-generated pools using the $K$ best-performing retrieval models.
The best-performing model is selected according to the ranking based on using a human-generated pool.
The LLM-generated pool is generated using the best pool generation model from Table~\ref{tab:1-shot-rank}, i.e., one-shot labeling ChatGPT using temperature of 0. Considering the 4 best-performing retrieval models, the relative ranking using the LLM-generated assessments is the same as using the human-generated assessments over all ranking metrics.
Using the LLM-generated assessments, the relative ranking of the 10 best-performing runs based on ndcg\_cut\_5 is the same as the relative ranking of these runs based on human-generated assessments (Kendall's Tau $=1$). 

According to Figure~\ref{fig:kendall@k}, as the value of $K$ increases (more runs are included in the comparison), the value of the correlation converges. This finding represents the reliability of LLM-generated assessments in terms of relative ranking.

\subsection{Filling judgment holes (RQ3)}
\subsubsection{\bf Experimental design} 
To answer our \textit{\ref{RQ3}: How are new models with different levels of holes ranked using LLM-generated assessments? Can we rely on LLM-generated labels to compare a new model with existing models?}, we simulate the case where a new model is being tested against TREC iKAT 2023. To do so, we run multiple experiments where in each one we take out all the relevance judgments of one of the models while keeping those judgments that are in common with other existing models. This leads to different levels of holes per model, depending on their similarity to other existing models. We then assess the relevance of the unjudged passages using ChatGPT and use those labels to compute the performance of the model. To assess the performance, we compare the ranking of the model using the original human assessments vs.\ ChatGPT-generated assessments and report the absolute difference in the model's ranking in the two cases. For example, if a model is ranked 3 using the human-generated labels, and ranked 7 when we use ChatGPT-generated labels, then the absolute relative ranking difference is 4, therefore the lower its value the better. This indicates, how reliable LLM-generated labels are in filling the holes for new models. 
The TREC iKAT 2023 includes 28 runs and baselines for ranking. The pooling is done based on the top 10 passages returned by each of these runs. 
After removing a run and generating labels for it using ChatGPT, 
we do the ranking based on the (1) new pool (human pool filled by LLM) and (2) human pool which includes the human judgments for the holes of the current run.

\subsubsection{\bf Results}
\begin{figure}
    \centering
    \includegraphics[width=0.5\textwidth]{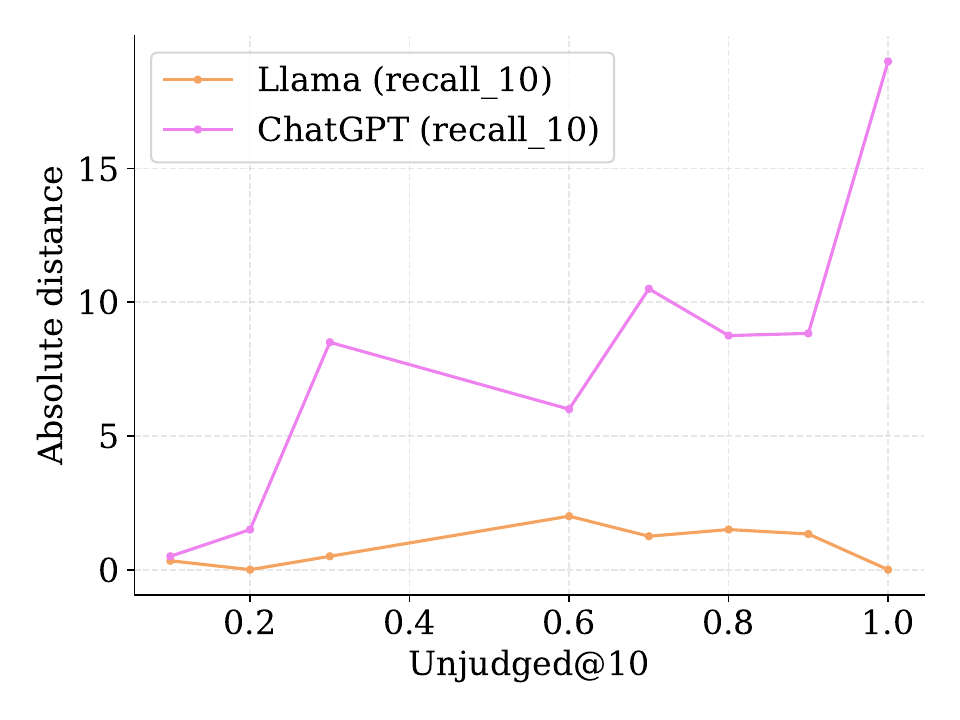}
    \caption{Absolute distance between the location of a new run before and after filling the holes using ChatGPT and LLaMA. 
The X-axis shows the average portion of unjudged documents among the top 10 documents returned by the run in the existing human-generated pool. We use the one-shot ChatGPT model with a temperature of 0 and zero-shot LLaMA-3-inst for hole filling. }
    \label{fig:hole_dist}
\end{figure}

The value of the absolute distance in the two rankings based on the portion of the Unjudged@10 passages for that specific run is shown in Figure~\ref{fig:hole_dist}. 
We use the one-shot ChatGPT model with a temperature of 0 for hole filling. 
As can be seen, as the value of Unjudged@10 increases, the absolute distance increases which means the missing run also increases. This makes sense because we know that ChatGPT is biased to rate the passages with higher scores compared to humans. 
As a result, we can conclude that given a new ranking model with a lot of missing judgments (a larger value for Unjudged@10), it is advisable to recreate the whole pool using the ChatGPT, rather than augmenting the existing human-created pool by filling the holes using ChatGPT.

Interestingly, we see that the results of LLaMA are exhibiting a completely different trend where the number of holes does not seem to matter. We see in the plot that LLaMA is able to consistently rank the missing run close to its original ranking and even achieves perfect ranking at some points. This is in line with our observation in Table~\ref{tab:kappa-test}, where we observed a higher agreement of LLaMA-generated labels with human labels, leading to a lower disparity in terms of the absolute value of the labels, which then makes the augmented labels more reliable.

%% file: Tables/confusion.tex
\begin{table}[t]
    \centering
    \caption{Graded-level confusion matrix over the test set using different fill holing models.}
    \begin{tabular}{lccccccc}
    \toprule
Hole Filling Model &    & 0 & 1 & 2 & 3 & 4 & sum \\ \midrule
Llama-3-inst FT    &  0 &  359  &  22 &  42 &  5   &  1  &  429   \\ 
                   &  1 &   2   &  1  &  9  &  2   &  0  &  ${14}^{*}$  \\ 
                   &  2 &   22  &  30 &  198&  52  &  4  &  306  \\ 
                   &  3 &   6   &  7  &  42 &  79  &  11 &  145   \\ 
                   &  4 &   0   &  0  &  4  &  12  &  7  &  ${23}^{*}$  \\ \midrule
                   
ChatGPT one-shot ($tmp=0$) &  0 &   221 &  7  &  21  &  9  &  0  &  258   \\ 
                         &  1 &   71  &  20 &  37  &  8  &  4  &  ${142}^{*}$   \\ 
                         &  2 &   42  &  8  &  57  &  22 &  4  &  133  \\ 
                         &  3 &   32  &  17 &  75  &  37 &  3  &  164  \\ 
                         &  4 &   23  &  8  &  105 &  74 &  12 &  ${222}^{*}$   \\      
      
\bottomrule
    \end{tabular}    
    \label{tab:confusion-graded}
\end{table}

\begin{table}[t]
    \centering
    \caption{Binary-level confusion matrix over the test set using different fill holing models. Scores higher than and equal to 2 are considered as relevant and scores lower than 2 are considered non-relevant.}
    \begin{tabular}{lcccc}
    \toprule
Hole Filling Model        &    & 0 & 1 & sum  \\ 
    \midrule
Llama-3 FT                &  0 & 384  &  ${59}^{*}$  & 443    \\ 
                          &  1 & ${65}^{*}$  &  409   & 474   \\  \midrule

ChatGPT one-shot (tmp=0)  &  0 &   319  &  ${79}^{*}$ &398    \\ 
                          &  1 &   ${130}^{*}$  &  389  &  519 \\  \midrule  

Llama-1 (pre-trained)~\citep{meng2024query}&  0 & 287   & 119  &  406 \\
                          &  1 & 162  & 349   & 511 \\                             
      
\bottomrule
    \end{tabular}
    \label{tab:confusion-binary}
\end{table}

%% file: Tables/kappa.tex
\begin{table}[!b]
    \centering
    \caption{Cohen's Kappa agreement between human and LLM labels on all the query-passage pairs from TREC iKAT 2023 dataset.}
    \begin{tabular}{lcc}
    \toprule
Hole Filling Model & Binary & Graded \\ 
\midrule

ChatGPT zero-shot  ($tmp=1$)& 0.170 & 0.099 \\ 
ChatGPT zero-shot ($tmp=0$) & 0.207 & 0.041 \\  \midrule

ChatGPT one-shot  ($tmp=1$)& 0.235 & 0.137 \\ 
ChatGPT one-shot  ($tmp=0$) & \textbf{0.269} & \textbf{0.155} \\ \midrule

ChatGPT two-shot  ($tmp=1$)& 0.133 & 0.076 \\ 
ChatGPT two-shot  ($tmp=0$) & 0.218 & 0.152 \\ \midrule

Llama-1 (pre-trained)~\citep{meng2024query} & 0.186 & - \\
\bottomrule
    \end{tabular}    
    \label{tab:kappa-all}
\end{table}

%% file: Tables/one-shot-ranking.tex
\begin{table*}[t]
    \centering
    \caption{Comparison between the relative ranking of TREC iKAT 2023 runs using LLM- and human-generated pools. In this table, the ChatGPT model is used as LLM, and the human-generated pool is recreated completely using different variations of prompts. The relative ranking is compared using Kendall's Tau ($\tau$), Spearman's Rho ($\rho$), and Rank-Biased Overlap (RBO) metrics.}
    
    \begin{tabular}{llcccccccc}
\toprule
Metric & Hole Filling Model  & ndcg\_cut\_3 & ndcg\_cut\_5 & ndcg & P\_10 & recall\_10 & recall\_1000 & map & recip\_rank \\ \midrule    
$\tau$   & zero-shot ($tmp=0$) & \textbf{0.862} & \textbf{0.862} & 0.841 & 0.836 & 0.794 & 0.709 & 0.730 & \textbf{0.873} \\ 
        & zero-shot ($tmp=1$) & 0.852          & \textbf{0.862} & 0.836 & 0.847 & 0.825 & 0.847 & 0.794 & 0.831 \\ \cmidrule{2-10} 

        & one-shot ($tmp=0$)  & 0.836          & 0.852 & \textbf{0.926} & 0.815 & 0.820 & \textbf{0.905} & \textbf{0.878} & 0.804 \\ 
        & one-shot ($tmp=1$)  & \textbf{0.862} & \textbf{0.862} & 0.899 & 0.836 & \textbf{0.836} & 0.868 & 0.852 & 0.804 \\ \cmidrule{2-10}   

       & two-shot ($tmp=0$)  & 0.831          & 0.825 & 0.852 & \textbf{0.868} & 0.831 & 0.889 & 0.836 & 0.862 \\ 
        & two-shot ($tmp=1$)  & 0.804          & 0.804 & 0.778 & 0.772 & 0.772 & 0.772 & 0.735 & 0.767 \\        

\bottomrule

$\rho$ & zero-shot ($tmp=0$) & 0.963 & \textbf{0.962} & 0.950 & 0.948 & 0.925 & 0.861 & 0.903 & 0.964 \\ 
       & zero-shot ($tmp=1$) & 0.959 & 0.961 & 0.947 & 0.959 & 0.942 & 0.959 & 0.921 & 0.952 \\ \cmidrule{2-10}

     & one-shot ($tmp=0$) & 0.955 & 0.960 & \textbf{0.979} & 0.938 & 0.949 & \textbf{0.980} & \textbf{0.960} & 0.936 \\ 
     & one-shot ($tmp=1$) & \textbf{0.964} & \textbf{0.962} & 0.973 & 0.948 & \textbf{0.956} & 0.969 & 0.945 & 0.935 \\ \cmidrule{2-10}
 
   & two-shot ($tmp=0$) & 0.949 & 0.952 & 0.947 & \textbf{0.968} & 0.953 & 0.975 & 0.944 & \textbf{0.967} \\ 
   & two-shot ($tmp=1$)& 0.942 & 0.946 & 0.921 & 0.918 & 0.913 & 0.909 & 0.887 & 0.918 \\

\bottomrule

RBO     & zero-shot ($tmp=0$) & 0.906 & 0.911 & 0.873 & 0.936 & 0.920 & 0.743 & 0.878 & 0.909 \\   
        & zero-shot ($tmp=1$) & 0.894 & 0.909 & 0.885 & 0.939 & \textbf{0.939} & 0.838 & 0.917 & 0.883 \\ \cmidrule{2-10}

        & one-shot ($tmp=0$)  & 0.927 & 0.941 & \textbf{0.955} & 0.924 & 0.900 & \textbf{0.893} & \textbf{0.962} & \textbf{0.926} \\ 
        & one-shot ($tmp=1$)  & \textbf{0.948} & \textbf{0.952} & 0.949 & 0.934 & 0.905 & 0.862 & 0.951 & 0.910 \\ \cmidrule{2-10}
        
        & two-shot ($tmp=0$)  & 0.897 & 0.901 & 0.940 & \textbf{0.941} & 0.909 & 0.880 & 0.937 & 0.904 \\ 
        & two-shot ($tmp=1$)  & 0.909 & 0.873 & 0.863 & 0.910 & 0.872 & 0.787 & 0.883 & 0.818 \\ 
        
\bottomrule

    \end{tabular}
    \label{tab:1-shot-rank}
\end{table*}

%% file: sections/conclusion.tex
\section{Conclusion}
In this work, we conducted extensive experiments to study the effect of LLM-generated relevance judgments on incomplete relevance judgments (aka.\ ``holes'') of the TREC iKAT 2023 collection. We studied the effectiveness of different open-source and closed-source on generating relevance assessments on the same set, where we observed that LLaMA-based labels align better with human labels.  
In line with previous work, we observed that automatic judgments from LLMs result in highly correlated model rankings; however, we found substantially lower correlates when human plus automatic judgments are used (regardless of LLM, one/two/few shot, or fine-tuned) when a new model was being assessed on the pool. We further found that, depending on the LLM employed, new runs will be highly favored (or penalized), and this effect is magnified proportional to the size of the holes. We conclude that generating automatic labels on the whole pool is more effective, rather than just the missing holes, as it leads to higher correlation and ensures that the same labeling biases are applied to all the models. 
Future work is required to prompt engineering and fine-tuning LLMs to reflect and represent the human annotations, in order to ground and align the models, such that they are more fit for purpose. Moreover, we plan to simulate various labeling strategies to study the effectiveness of fine-tuning in more practical scenarios.